Intrinsic pinning property of FeSe$_{0.5}$Te$_{0.5}$

M. Migita[1,*], Y. Takikawa[1], M. Takeda[1], M. Uehara[1], T. Kuramoto[1], Y. Takano[2,3], Y. Mizuguchi[2,3], Y. Kimishima[1]

[1]Dept. of Phys., Fac. of Eng., Graduate School of Yokohama Nat'l Univ., Tokiwadai 79-5, Hodogaya, Yokohama 240-8501, Japan

[2]National Institute for Materials Science, 1-2-1, Sengen, Tsukuba 305-0047, Japan

[3]Japan Science and Technology Agency, Transformative Research-project on Iron Pnictides

**Abstract**

The intrinsic pinning properties of FeSe$_{0.5}$Te$_{0.5}$, which is the superconductor with $T_c$ of about 14 K, were studied by the analysis of magnetization curves by the extended critical state model. In the magnetization measurements by SQUID magnetometer, the external magnetic fields were applied parallel and perpendicular to $c$-axis of the sample. The critical current density $J_c$'s under the perpendicular field of 1 T were estimated by using the Kimishima model as about $1.6 \times 10^4$, $8.8 \times 10^3$, $4.1 \times 10^3$, and $1.5 \times 10^3$ A/cm$^2$ at 5, 7, 9, and 11 K, respectively, and the temperature dependence of $J_c$ could be fitted with the exponential law of $J_c(0) \times \exp(-\alpha T / T_c)$ up to 9 K and power law of $J_c(0) \times (1 - T / T_c)^n$ near $T_c$.

*Corresponding author

Mr. M. Migita

Postal address : Kimishima Lab., Department of Physics, Graduated School of Engineering, Yokohama National University, Tokiwadai 79-5, Hodogaya-ku, Yokohama 240-8501, Japan

Phone : +81-45-339-4182;

Fax : +81-45-339-4182;

E-mail address : d10sd203@ynu.ac.jp

# 1. Introduction

Since the discovery of high temperature superconductivity in LaFeAsO$_{1-x}$F$_x$ with $T_c$ of 26 K [1], the research of iron-based superconductivity has been performed by many researchers. Particularly the FeSe with $T_c$ of 8 K has the simplest structure among the other iron-based superconductors [2]. The $T_c$ of FeSe increased up to about 15 K by partial substitution of Te or S for Se [3-8]. This system is advantageous to industrial application because tellurium, selenium, and sulfur have low toxicity in comparison with the arsenic. Furthermore, it shows the superconductivity below the *refrigerator temperature* of 20 K.

Here, we report the intrinsic pinning properties of FeSe$_{0.5}$Te$_{0.5}$, which is the superconductor with $T_c$ of about 14 K, were studied by the analysis of magnetization curves by the extended critical state model by Kimishima [9-11].

# 2. Sample preparation

The FeSe$_{0.5}$Te$_{0.5}$ sample was prepared by solid-state reaction. Powders of Fe, Se, Te were ground and mixed with the nominal stoichiometry. The fully mixed powder were put into the alumina tube and then sealed in an evacuated quartz tube, which was heated at 1323 K for 20 hr. Subsequently it was cooled to 673 K and annealed for 200 hr to stabilize the superconductive layers. The prepared sample with the size of 1.5×1.0×0.5 mm$^3$ was characterized by powder X-ray diffraction (XRD) using Cu-$K\alpha$ radiation. Temperature dependence of magnetic susceptibility and external magnetic field dependence of magnetization were measured using MPMS SQUID (superconducting quantum interference device) magneto-meter by Quantum Design.

## 3. Results and discussions

The XRD pattern is shown in Fig. 1. Most of the peaks were well indexed using a space group of *P4/nmm* for tetragonal FeSe$_{0.5}$Te$_{0.5}$ superconducting phase. The calculated lattice parameters were $a$ = 0.37963 nm, and $c$ = 0.60387 nm. These values are in good agreement with other reports [12, 13]. The small additive peaks of hexagonal FeSe$_{0.5}$Te$_{0.5}$ and iron, these were not superconductor, were observed as an impurity phase. Scanning electron microscope (SEM) image of *c*-plane is shown in Fig. 2, where the stratified crystal with the size of about 100 μm was observed.

Zero-field cooled (ZFC) and field cooled (FC) DC magnetic susceptibilities at 1 mT are shown in Fig. 3. The external magnetic fields $H$ were applied parallel and perpendicular to *c*-axis of the sample. Clear superconducting diamagnetism was observed below $T_c$ of 14 K. Small ZFC signals under the magnetic field perpendicular to

*c*-axis means that the magnetic field penetrates into the space between the superconducting layers, and the effective superconducting volume becomes small. Figs. 4(a) and (b) show the magnetization hysteresis loops under the magnetic field parallel and perpendicular to *c*-axis, respectively, at T = 5, 7, 9, 11, and 13 K between the field of -5 T and 5 T. Here the ferromagnetic magnetizations of iron impurity at T = 15 K were subtracted. It may be due to the small interlayer $J_c$ that the magnitude of hysteresis under the field parallel to *c*-axis is larger than that under the field perpendicular to *c*-axis. An indication of "fish-tail" like humps was observed in the *B*//*c*-axis magnetizations at 5 K, and 7 K, which are similar to YBCO crystals [14].

In this report, we analyze $J_c$ from $B \perp c$-axis magnetization with extended critical state model by Kimishima [9-11] based on the well-known Kim model [15] and Bean model [16]. In this model the value of the internal critical current density $J_c$ was estimated by following equation

$$J_c(H) = \frac{B_{eq}^*(B_{eq}^* + 2B_0)}{2\mu_0 a\{|B_{eq}(H)| + B_0\}} \tag{1}.$$

In this equation, $2a$ is the thickness of slab sample. $B_{eq}(H)$ $(= \mu_0[H + M_{eq}(H)])$ is the flux density at inner sample surface, where $M_{eq}$ is the equilibrium magnetization. $B_{eq}^* = B_{eq}(H_p)$, and $H_p$ is the full penetration field above which the internal flux density becomes non-zero at the center of the slab. $B_0$ can be called as the cross-over flux density between the Anderson-Kim equations [15] and the Bean-model [16]. In Fig. 5, the experimental and theoretical hysteresis loops of magnetizations were depicted between -1 T and 1 T in the case of $B \perp c$-axis. In this external magnetic field area, experimental values were well fitted by the theoretical results. Fitting parameter $B_0$ is larger enough than $B_{eq}^{*}$ in the eq.(1) at low temperature. Therefore, it is thought that $J_c$ in the FeSe$_{0.5}$Te$_{0.5}$ sample has relatively uniform distribution as in the Bean state at low temperature.

Fig. 6 shows the theoretical $J_c(B)$ curves in the case of $B \perp c$-axis, where the sample thickness of 100 μm by the SEM image was used for the theoretical calculation in eq.(1). The $J_c$ at 1 T was estimated as about $1.6 \times 10^4$, $8.8 \times 10^3$, $4.1 \times 10^3$, and $1.5 \times 10^3$ A/cm$^2$ at 5, 7, 9, and 11 K, respectively.

Temperature dependence of $J_c$ at 0 T and 1 T were shown in Fig. 7. Feigel'man *et al* [17] insist the current relaxation due to thermal activation process for the single vortex creep was expressed by exponential law as

$$J_c = J_c(0) \exp[-\alpha T] \tag{2}.$$

The parameter $\alpha$ in eq.(2) was given by $[U_c \ln(t/\tau_0)]^{-1}$, where $U_c$ is the energy barrier for the single vortex pinning, and $\tau_0$ is the relaxation time of the single vortex creep. The experimental data obeyed the exponential law between 5 K and 9 K (Fig. 7. solid line), where the fitting parameters are $J_c(0) = 1.81 \times 10^5$ A/cm$^2$, $\alpha = 0.19$ at 0 T, and $J_c(0) = 9.44 \times 10^4$ A/cm$^2$, $\alpha = 0.35$ at 1 T. It shows that the single vortex phase exists up to 9 K. On the other hand, the experimental data obeys the following power law near $T_c$,

$$J_c = J_c(0) \times (1 - T / T_c)^n \tag{3},$$

which generally holds for the sintered superconductors near $T_c$ [18]. As for the presented data, $J_c(0) = 1.67 \times 10^5$ A/cm$^2$, $n = 1.6$ at 0 T, and $J_c(0) = 3.64 \times 10^4$ A/cm$^2$, $n = 2.1$ at 1 T (Fig. 7. dashed line). The critical exponent $n$ has been predicted to be 1 for the Josephson tunnel junctions [19] and 3/2 for the micro-bridge junctions by Ginzburg-Landau theory [20]. Our results of $n$'s at 0 T may show the formation of micro-bridge junctions near $T_c$ as in the High $T_c$ materials [21, 22].

## 4. Conclusion

The critical current density $J_c$'s in FeSe$_{0.5}$Te$_{0.5}$ under the field perpendicular to $c$-axis were estimated by using the Kimishima model. The $J_c$ at 1 T was estimated as about 1.6 × 10$^4$, 8.8 × 10$^3$, 4.1 × 10$^3$, and 1.5 × 10$^3$ A/cm$^2$ at 5, 7, 9, and 11 K, respectively. Temperature dependence of $J_c$ at 0 T and 1 T were obeyed the exponential law of eq.(2) between 5 K and 9 K. It shows that the single vortex phase exists up to 9 K. On the other hand, the experimental data obeys the power law of eq.(3) near $T_c$. The fitting parameter of $n$'s value of 0 T is 1.6 which may show the formation of micro-bridge junctions near $T_c$ as in the High $T_c$ materials.

## Acknowledgment

The FeSe$_{0.5}$Te$_{0.5}$ sample preparation was supported by *National Institute for Materials Science* (NIMS).

**Figure captions**

Fig. 1 : X-ray diffraction pattern of $FeSe_{0.5}Te_{0.5}$.

Fig. 2 : SEM image of $c$-plane of $FeSe_{0.5}Te_{0.5}$.

Fig. 3 : Zero-field cooled (ZFC) and field cooled (FC) DC magnetic susceptibilities at 1 mT with $B//c$-axis and $B\perp c$-axis.

Fig. 4 : Hysteresis loops of $B//c$-axis magnetization (a) and $B\perp c$-axis one (b) at T = 5, 7, 9, 11, and 13 K, where the ferromagnetic magnetizations of iron impurity at T = 15 K were subtracted.

Fig. 5 : Experimental and theoretical hysteresis loops with $B\perp c$-axis between -1 T and 1 T, where the ferromagnetic magnetizations of iron impurity at T = 15 K were subtracted.

Fig. 6 : Theoretical $J_c(B)$ curves with $B\perp c$-axis.

Fig. 7 : Temperature dependence of $J_c$ at 0 T and 1 T from $B\perp c$-axis magnetization.

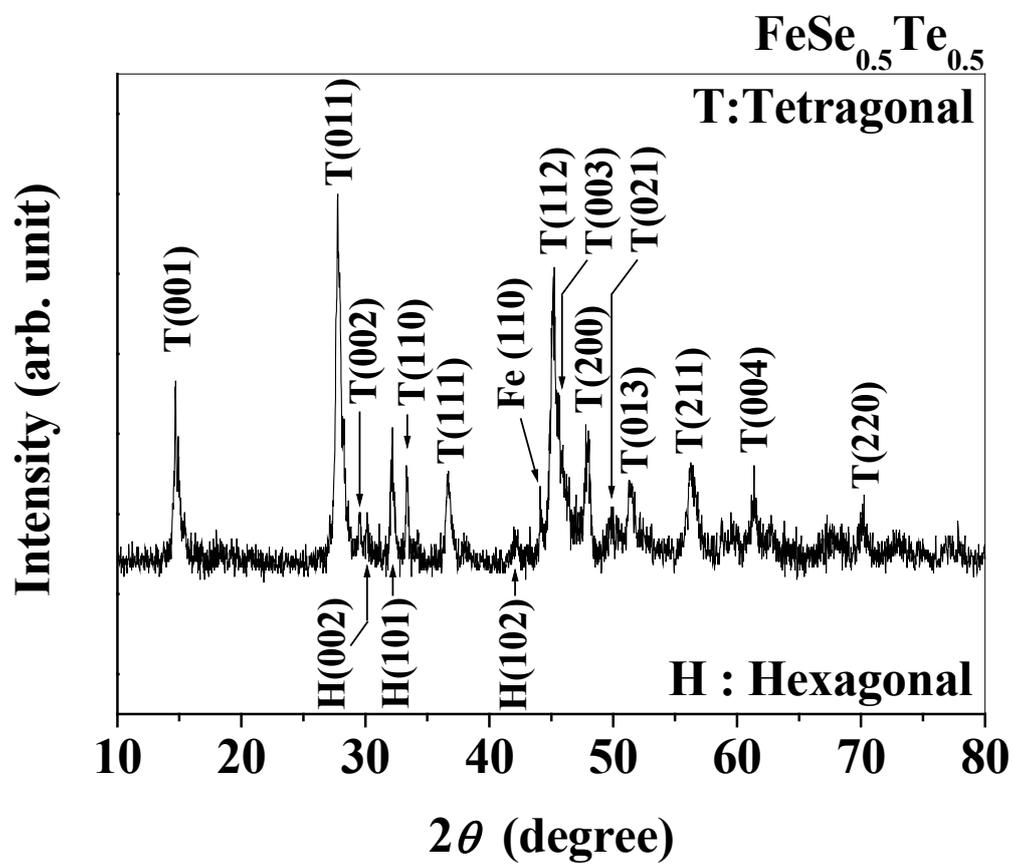

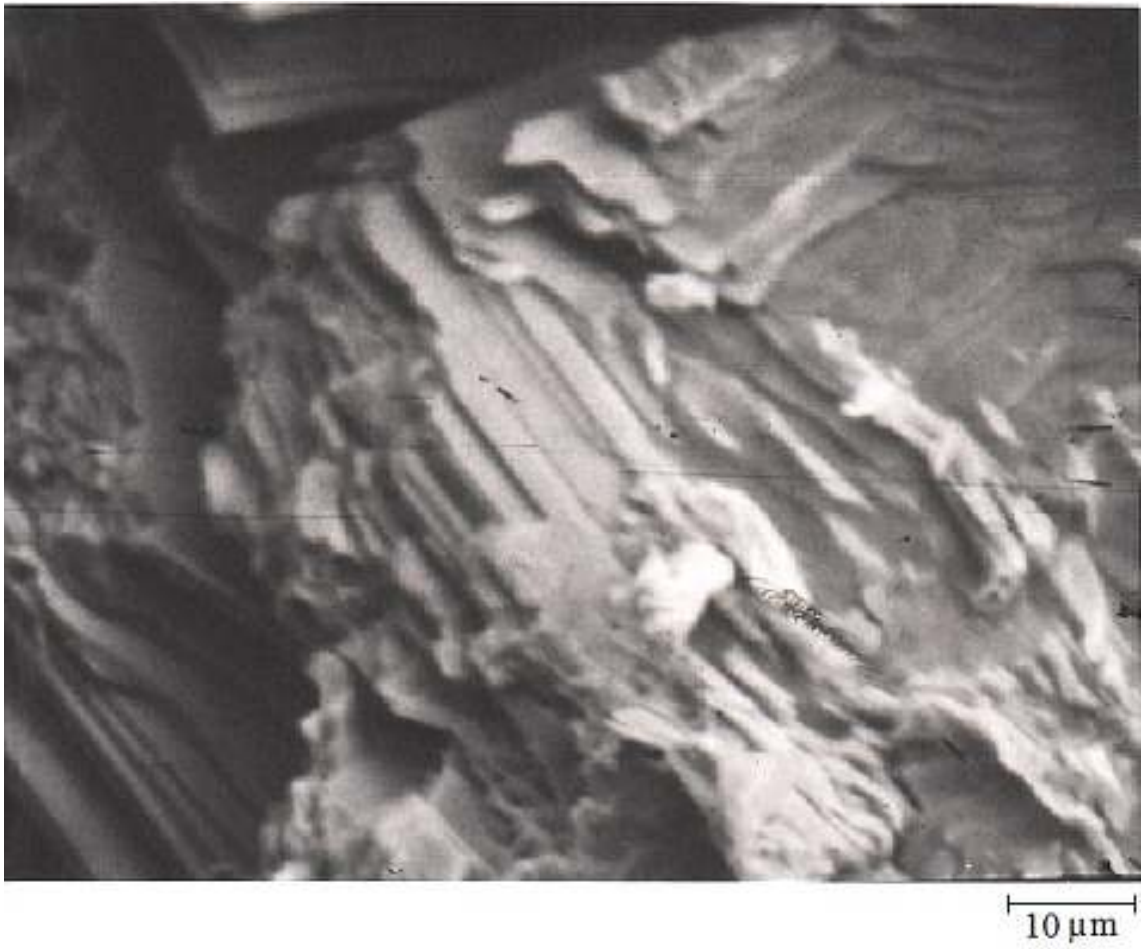

10 µm

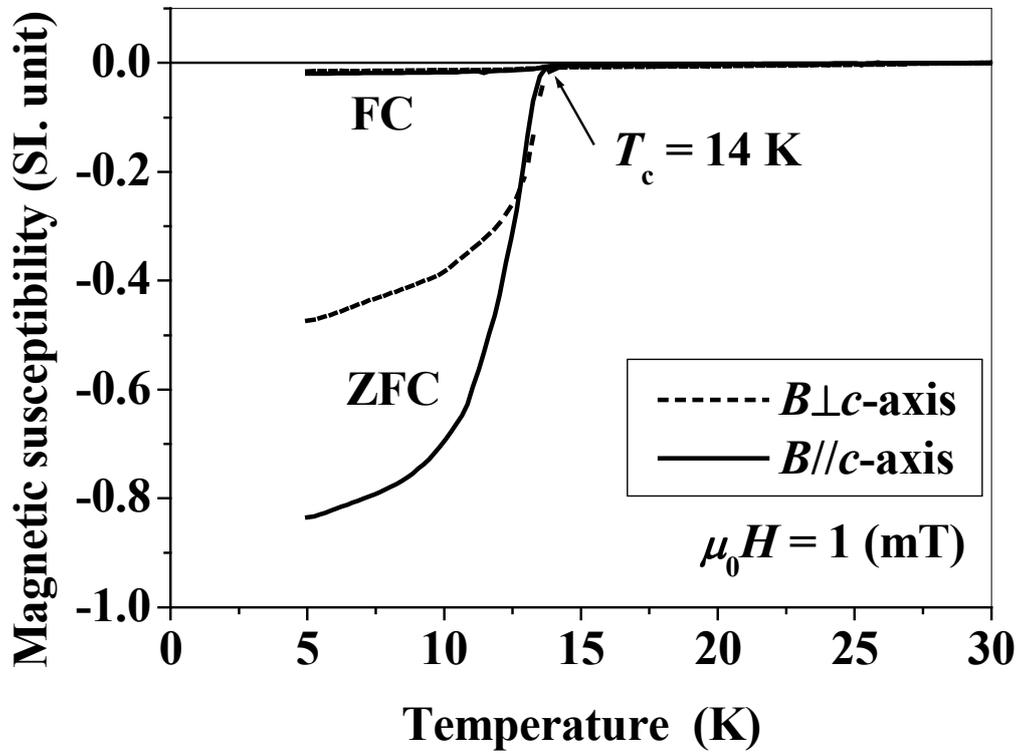

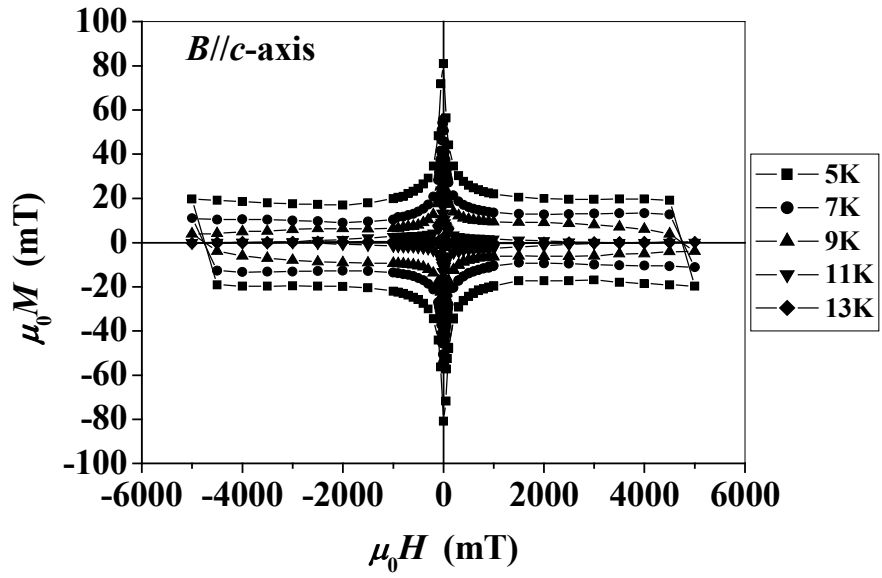

**(a)**

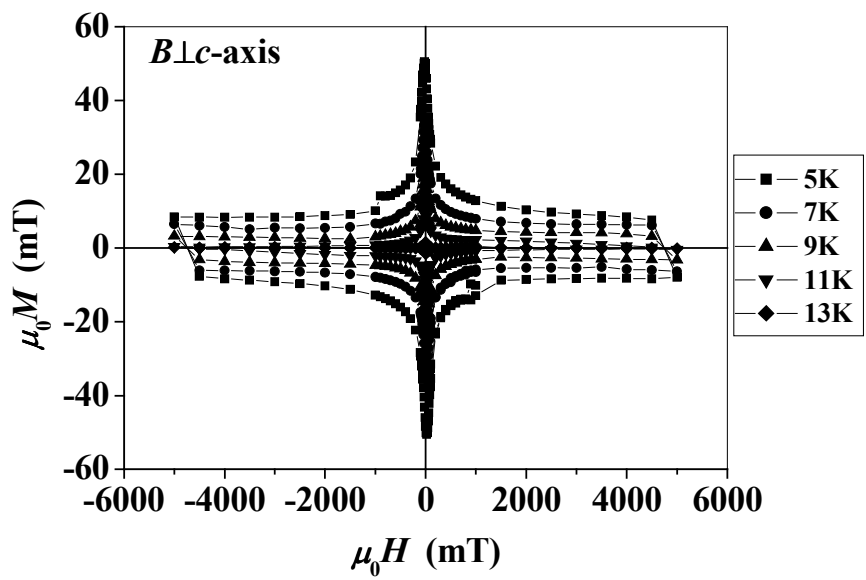

**(b)**

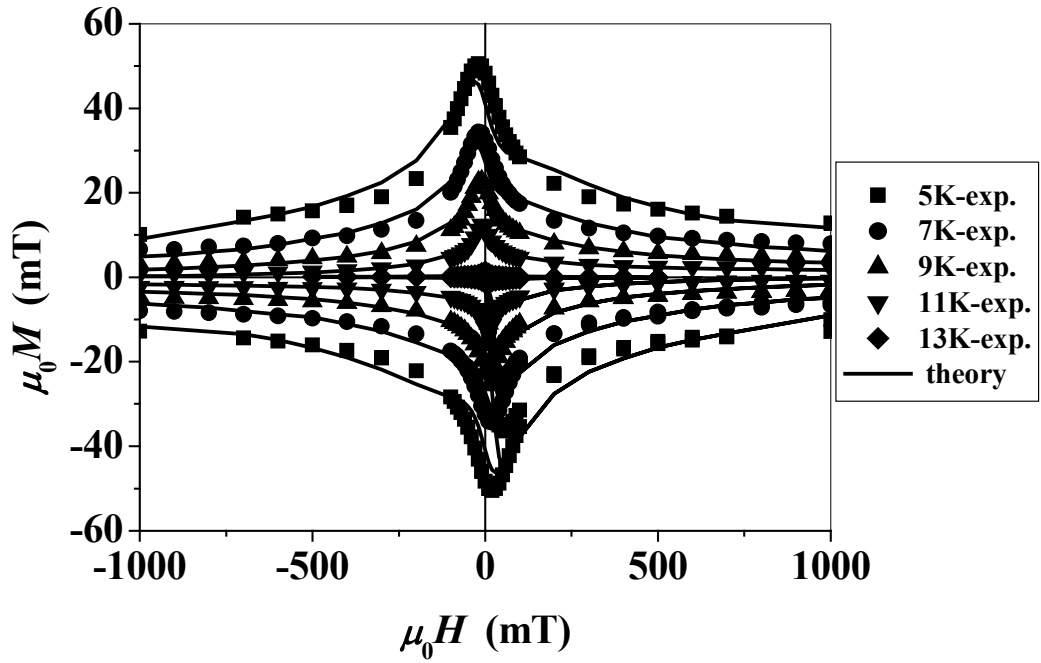

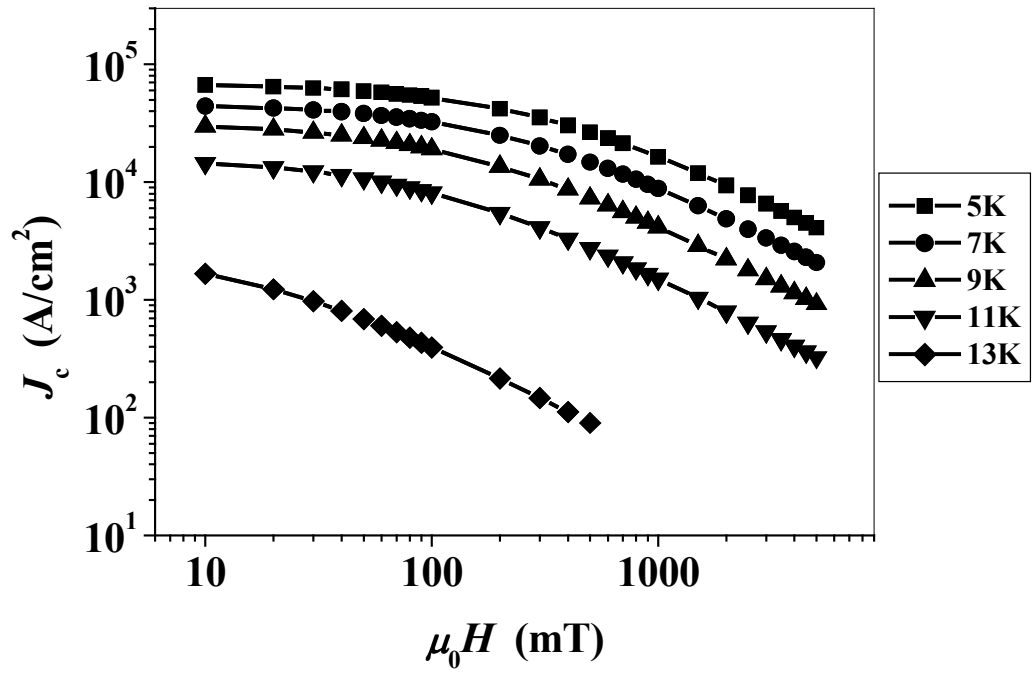

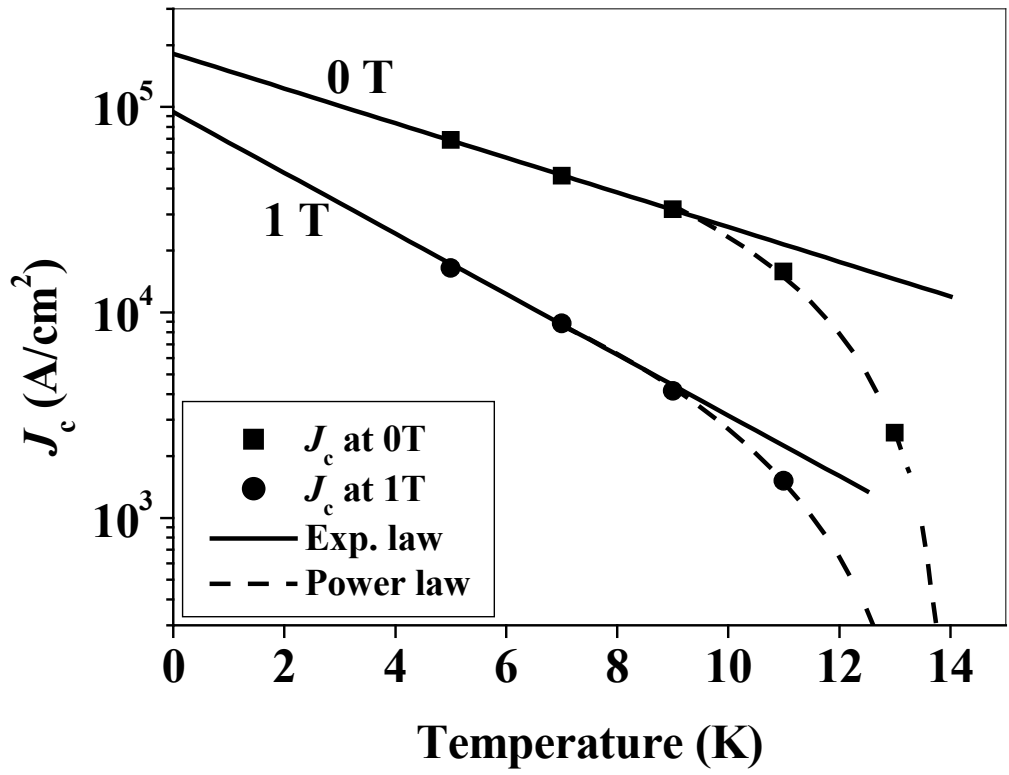